# Comparative Study of Nonproportionality and Electronic Band Structures Features in Scintillator Materials

Wahyu Setyawan, Romain M. Gaume, Robert S. Feigelson, and Stefano Curtarolo

*Abstract*—The origin of nonproportionality in scintillator materials has been a long standing problem for more than four decades. In this manuscript, we show that, with the help of first principle modeling, the parameterization of the nonproportionality for several systems, with respect to their band structure curvature suggests a correlation between carrier effective mass and energy response. We attribute this correlation to the case where free electrons and holes are the major energy carriers. Excitonic scintillators do not show such a definitive trend. This model suggests a potential *high-throughput* approach for discovering novel proportional scintillators in the former class of materials.

*Index Terms*—Electronic Structures, Nonproportionality, Photon Response, Scintillator Materials.

## I. Introduction

SCINTILLATION detectors have demonstrated great potential for nuclear and radiological surveillance. Indeed, scintillation-based detectors offer numerous advantages including high light yields, low attenuation lengths, fast responses, large active areas, and room-temperature operation [1, 2]. The optimization of these parameters largely contributes to improving detector efficiency [3]. The ability of a detector system to distinguish between radiation from threatening and non-threatening sources depends on its capability to unambiguously resolve radioisotope-decay energy signatures. This ability not only requires the appropriate energy attribution of detector response to gamma-rays, but also sufficient energy resolution to discriminate between the energy fingerprints in the gamma-ray spectrum [4]. The performance of a scintillation detector system is determined by the sensitivity and the energy resolution of its components. In that regard, considerable effort has been invested to improve the resolution of scintillator materials and light detectors. As early as the 1960s, it was recognized that the intrinsic resolution of the scintillator material was the dominant contributor to the overall energy resolution of a detector system [5-8]. The intrinsic energy resolution of the Tl:NaI has been reported to contribute up to 86 % of the overall detector resolution, while the resolution of the light detector contributes only 4 % [5, 9, 10], thus the resolution of the scintillator material continues to be a primary factor limiting scintillation detector performance.

In the search for highly efficient scintillator materials, the nonproportional relationship between the energy of the detected radiation and the light yield has been identified as a crucial issue contributing to intrinsic energy resolution [4, 5]. This nonproportional behavior was first observed in 1950 by Pringle and Standil in Tl:NaI [11]. This discovery motivated an effort to identify the physical phenomena associated with nonproportionality. The first theoretical interpretation of these observed behaviors were proposed by Taylor *et al.* in 1951 [12]. The authors analyzed the consequences on light output of a lossy electron production cascade that depends on the incident radiation energy. Their model was then extended by Murray and Meyer (1961) [13] and Hill and Collinson (1966) [14, 15]. They respectively clarified the role played by exciton recombination at activator ion sites and light emission re-absorption on nonproportionality. In particular, they showed that the decrease in response of NaI and CsI to heavy charged particles with increasing particle's mass [13] was strongly correlated to differences in ionization density. These early models also highlighted the two main processes during which energy losses that account for nonproportional behavior occur: those during the absorption and the electron cascade related to the host material and those associated with exciton transport to the activator ions.

Rising interest in the development of more efficient scintillation detectors has precipitated experimental efforts concentrated on understanding the effect of intrinsic energy resolution in a broader range of materials over the last two decades [4]. These systematic studies of nonproportionality have been contingent upon the availability of a set of radioisotope sources covering a broad spectrum of energies. Fortunately, the development of the Compton Coincidence

Manuscript received December 22, 2008. This work was supported in part by the U.S. Department of Homeland Security – Domestic Nuclear Detection Office.

Wahyu Setyawan, is with Department of Mechanical Engineering and Materials Science, Duke University, Durham, NC 27708 USA (e-mail: wahyu.setyawan@duke.edu).

Romain M. Gaume, is with Department of Materials Science, Stanford University, Palo Alto, CA 94305 USA (e-mail: rgaume@stanford.edu).

Robert S. Feigelson, is with Department of Materials Science, Stanford University, Palo Alto, CA 94305 USA (e-mail: feigel@stanford.edu).

Stefano Curtarolo is with Department of Mechanical Engineering and Materials Science, Duke University, Durham, NC 27708 USA (corresponding author, phone: 919-660-5506; fax: 919-660-8963; e-mail: stefano@duke.edu).

Technique (CCT) by Valentine in 1994 [16] has increased the ease with which the response of a range of materials may be investigated. During the last decade, efforts have been devoted to characterizing the energy response of a variety of recently-identified scintillator materials including lutetium orthosilicate (LSO), lutetium aluminum perovskite (LuAP), lanthanum bromide (LaBr$_3$), and lead tungstate (PWO) [4]. These measurements have been limited to simple compositional variations (e.g. cation substitution, dopant concentration) on a handful of monocrystalline scintillators. Concurrently, Monte Carlo simulations have been successfully employed to replicate some features of experimental studies in Tl:NaI [4, 10]. The simulations were used to determine the energy loss rate $dE/dx$ that enabled the calculations of the density of activated sites $n(\rho,x,t)$ and the local light yield produced by these activated sites using a kinetic model that accounts for spontaneous emission, quenching, and transfer rates [17].

A different research approach was undertaken to relate nonproportionality to relevant material parameters. Balcerzyk *et al.* suggested that there exists a correlation between crystal structure and energy response [18, 19]. For instance, similar response curves were observed in cerium-doped, rare-earth orthosilicates (LSO, GSO, YSO) [19] as well as in alkali iodides (Tl:NaI, Tl:CsI, Na:CsI) [20]. In addition, low concentrations of activator and impurity ions have been shown to affect the nonproportionality to a lesser extent than substantial cationic or anionic substitutions (e.g. Y$_{1-x}$Gd$_x$SO) [21]. However, this crystal structure-response relationship is still not fully understood, and further studies are required.

In this paper, we report for the first time the observation of possible correlation between the electronic structure of various scintillator materials and their energy response. We start from the ansatz that the unbalance between carrier mobilities relates to the "difficulty" of reaching recombination sites and emitting light photons. Depending on the nature of the carriers, we identify two classes of materials following different trends.

## II. METHOD

We have addressed the following inorganic scintillator materials:
   a) cubic: LiBaF$_3$ (LBF), BaHfO$_3$, SrHfO$_3$, CsI, NaI, BaF$_2$, Bi$_4$Ge$_3$O$_{12}$ (BGO), Y$_3$Al$_5$O$_{12}$ (YAG), Lu$_3$Al$_5$O$_{12}$ (LuAG), ZnSe
   b) orthorhombic: SrI$_2$, YAlO$_3$ (YAP), LuAlO$_3$ (LuAP), K$_2$LaCl$_5$ (KLC)
   c) hexagonal: LaBr$_3$, LaCl$_3$
   d) monoclinic: Y$_2$SiO$_5$ (YSO), Lu$_2$SiO$_5$ (LSO)

The electronic structures of these materials were calculated using the Vienna Ab-initio Simulation Package (VASP) [22] with projector augmented waves pseudopotentials [23] and exchange-correlation functionals as parameterized by Perdew-Burke-Ernzerhof [24] within the generalized gradient approximation. Calculations were performed at zero temperature and without zero-point motion. All structures were fully relaxed with energy convergence better than 1 meV/atom. To get an accurate density of states and charge densities for the band structure calculations, we used dense **k**-point meshes of about 27,500 per number of atoms in the unit cell. In this study, we also calculated the band structures at pressures $P = 0.5, 1, 2, 4, 6, 8,$ and 10 GPa by applying a Pulay stress. This stress, which arises from the fact that the plane wave basis set employed in VASP is incomplete with respect to volume change, can be used to relax a structure under isotropic external pressure [22]. In relaxing the structures to the equilibrium volume, we have used large plane wave cut-off to increase basis set completeness. Throughout the calculations, the uncertainty in the pressure due to the kinetic energy cutoff of the plane wave basis sets was always less than 50 MPa.

For lanthanum-containing compounds, we apply the GGA+U technique [25,26] to the La(4f) states to correct their position relative to La(5d) levels. We choose $U_{eff} = 7.5$ eV to reproduce the experimentally observed position of 4f bands [27].

The dispersion of the band structure was characterized by effective masses of electrons ($m_e$) and holes ($m_h$) calculated from the second derivative of energy with respect to **k**-vector at conduction band minimum (CBM) and valence band maximum (VBM), respectively. The CBM and VBM may contain more than one band (degenerate bands) and may occur at more than one **k**-point (degenerate optima). As a first approximation, $m_e$ and $m_h$ were calculated as averages:

$$m_e = \frac{1}{N} \sum_{i=1}^{N} m_{e,i} \tag{1}$$

$$m_h = \frac{1}{N} \sum_{i=1}^{N} m_{h,i} \tag{2}$$

The index $i$ runs over all the degenerate bands and all the degenerate optima. Note that the individual effective mass $m_{e,i}$ or $m_{h,i}$ is calculated from a symmetric optimum, e.g. in an orthorhombic cell, if the only optimum is at Γ-point, there will be three contributing curvatures corresponding to X-Γ-X, Y-Γ-Y, and Z-Γ-Z segments. In addition, to get an idea of the similarity of the effective masses that are experienced by electrons and holes, we define an effective mass ratio, $m_r$, as follows:

$$m_r = \max[m_h/m_e, \, m_e/m_h] \tag{3}$$

Using this definition, $m_r$ is always larger than or equal to one. The nonproportionality of the photon response is calculated as:

$$NP_{10/662} = Y_{10\,\text{keV}} / Y_{662\,\text{keV}} \tag{4}$$

$$\sigma_{NP} = \sqrt{\frac{1}{N} \sum_{i=1}^{N} \left(1 - \frac{Y_i}{Y_{662\,keV}}\right)^2} \tag{5}$$

where $Y_i$ denotes the photon light yield resulting from gamma radiation with energy $E_i$. To obtain $NP_{10/662}$ for materials exhibiting nonmonotonic responses, such as NaI, CsI, SrI$_2$, and BGO, we take, in the response spectrum $Y(E)/Y_{662}$, the most deviated point from 1 among all data with energies larger than 10 keV. The quantity $\sigma_{NP}$, measuring the nonproportionality, is calculated from ten data points evenly spaced in a logarithmic scale from 10 keV $\leq E_i <$ 1 MeV as previously suggested [28].

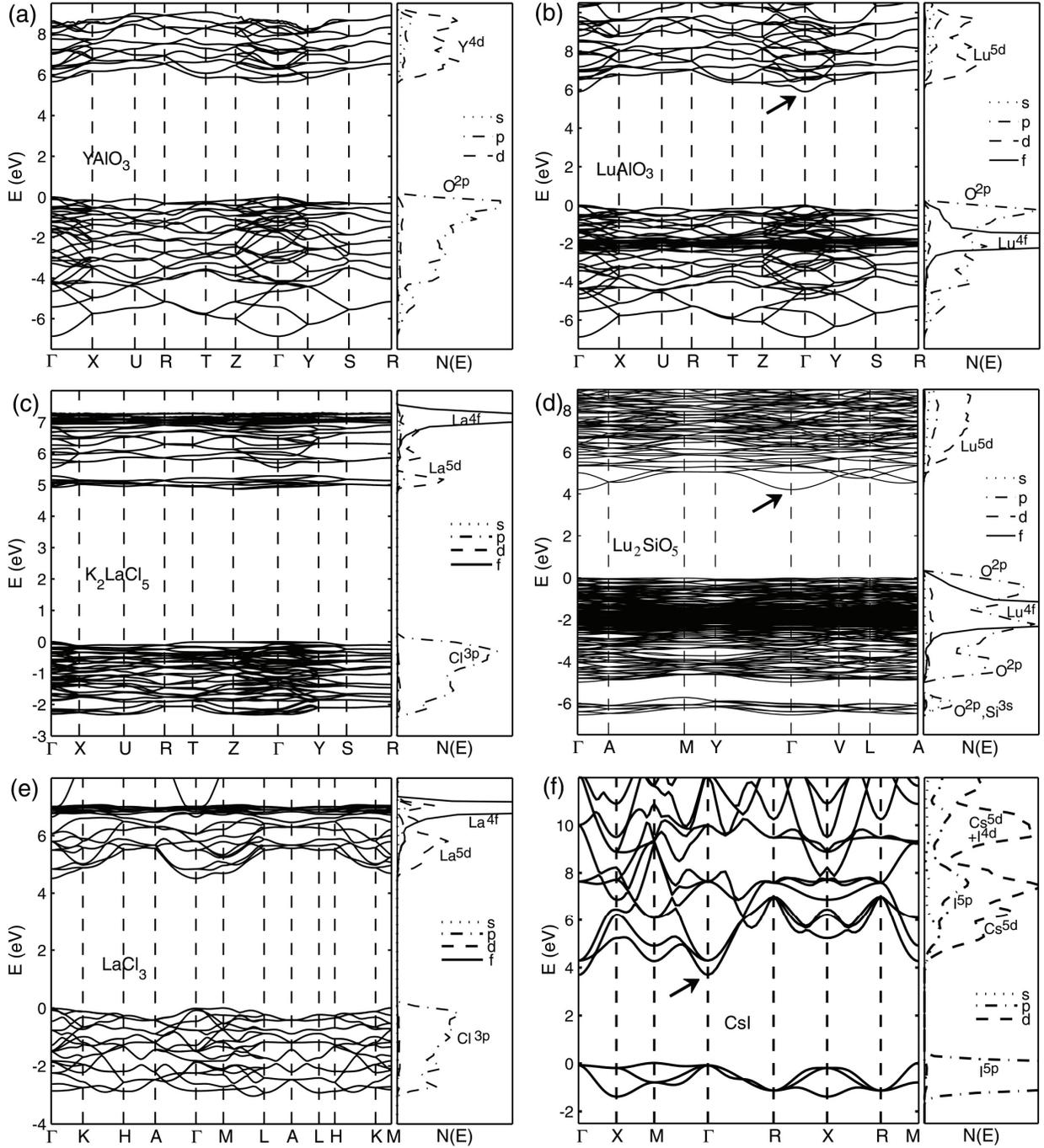

Fig. 1.  Calculated electronic structures of materials with a good proportional response (left panels) and a poor proportionality (right panels). The scale is shifted so that the top of valence bands is at zero. Note the dip in the bands of materials with poor proportionality as indicated with arrows.

## III. RESULTS

Figure 1 shows the valence and conduction bands of selected crystals. Materials in the left panel (YAP, KLC, LaCl₃) have a good proportional response down to 10 keV.

Systems with poor proportionality are illustrated in the right panel (LuAP, LSO, CsI). For convenience, we shift the bands so that the top of the valence band levels at zero. The electronic density of state $N(E)$ is projected for each orbital to indicate the contribution to the bands. YAP and LuAP both crystallize in the same space group (Pnma, #62). The conduction bands in YAP and LuAP are dominated by the d-orbitals of the cations. Oxygen's 2p-orbitals are dominant in the valence bands with an addition of Lu's 4f-orbitals in the case of LuAP. Although both materials exhibit similar band dispersions, the bottom of the conduction band of YAP, located at gamma-point ($\Gamma$), is much flatter than the one of LuAP (indicated by the arrow). When we extend the comparison of the flatness of the band edges, we observe that materials in the right panel (LSO and CsI) experience a "dip" in the bottom of their conduction bands, whereas KLC and LaCl$_3$ (left panel) do not. This "dip", indicating a rapid change of curvature of a band in the momentum space, is a direct consequence of how atomic orbitals overlap each other in the solid. This in turn is expected to affect the transport mechanism and efficiency of charge carriers moving to the luminescence centers and hence the light yields.

In order to investigate any possible correlations between effective masses and nonproportionality, we have calculated $m_e$ and $m_h$ at ambient pressure using equations (1) and (2). For all the compounds under investigation, the effective mass data were compared to scintillation response measurements published in the literature by various authors. For this study, the electron response would be more appropriate than photon response [29]. However, because of the more limited data currently available in the literature for electron response, we are forced to use the photon response. The nonproportionality of the response is characterized by $NP_{10/662}$ and $\sigma_{NP}$ as defined in equations (4) and (5). In case of activated scintillators, since the activator concentrations are usually small, the calculated electronic structures were performed on the host undoped materials. Small concentrations of dopants are expected to shift the bands vertically by a small amount (rigid band model). The results are summarized in Table 1. Note that, for SrI$_2$, the nonproportionality is calculated from the electron response [30] since the photon response data are not available.

| Materials (spacegroup) | $m_e$ (k-point) [$m_0$] | $m_h$ (k-point) [$m_0$] | $NP_{10/662}$ | $\sigma_{NP}$ |
|---|---|---|---|---|
| Tl:NaI (#225) | 0.287 ($\Gamma$) | 2.397 ($\Gamma$) | 1.15[31] | 0.118 |
| Tl:CsI (#221) | 0.312 ($\Gamma$) | 2.270 (M) | 1.18[32] | 0.111 |
| Eu:SrI$_2$ (#61) | 0.288 ($\Gamma$) | 6.360 ($\Gamma$) | 1.04*[30] | 0.015* |
| Ce:K$_2$LaCl$_5$ (#62) | 2.207 (Z) | 10.61 ($\Gamma$) | 0.95[4] | 0.023 |
| Ce:LaBr$_3$ (#176) | 1.323 ($\Gamma$) | 2.325 ($\Gamma$) | 0.92[33] | 0.035 |
| Ce:LaCl$_3$ (#176) | 1.333 ($\Gamma$) | 2.191 (M) | 0.90[34] | 0.042 |
| BaF$_2$ (CV) (#225) | 1.729 (X) | 3.870 (X) | 0.80[4] | 0.078 |
| BaF$_2$ (#225) | 0.499 (X) | 4.190 ($\Sigma$) | - | - |
| Te:ZnSe (#216) | 0.146 ($\Gamma$) | 0.949 ($\Gamma$) | 0.83[35] | 0.076 |
| Ce:YAP (#62) | 2.335 (B) | 1.941 ($\Gamma$) | 0.95[18] | 0.025 |
| Ce:LuAP (#62) | 0.423 ($\Gamma$) | 2.094 ($\Gamma$) | 0.70[36] | 0.159 |
| Ce:YAG (#230) | 1.094 ($\Gamma$) | 1.975 ($\Gamma$) | 0.82[18] | 0.088 |
| Ce:LuAG (#230) | 0.979 ($\Gamma$) | 2.432 (H) | 0.72[18] | 0.136 |
| Ce:YSO (#15) | 0.699 ($\Gamma$) | 3.795 ($\Gamma$) | 0.65[19] | 0.204 |
| Ce:LSO (#15) | 0.545 ($\Gamma$) | 3.603 ($\Gamma$) | 0.62[19] | 0.216 |
| BGO (#220) | 0.599 ($\Delta$) | 3.022 (H) | 0.59[32] | 0.224 |
| Ce:LiBaF$_3$ (#221) | 0.610 ($\Gamma$) | 9.071 ($\Gamma$) | - | - |
| LiBaF$_3$ (#221) | 2.780 (R) | 4.675 (M) | - | - |
| Ce:BaHfO$_3$ (#221) | 0.647 ($\Gamma$) | 3.678 (M) | - | - |
| Ce:SrHfO$_3$ (#221) | 0.664 ($\Gamma$) | 2.394 (R) | - | - |

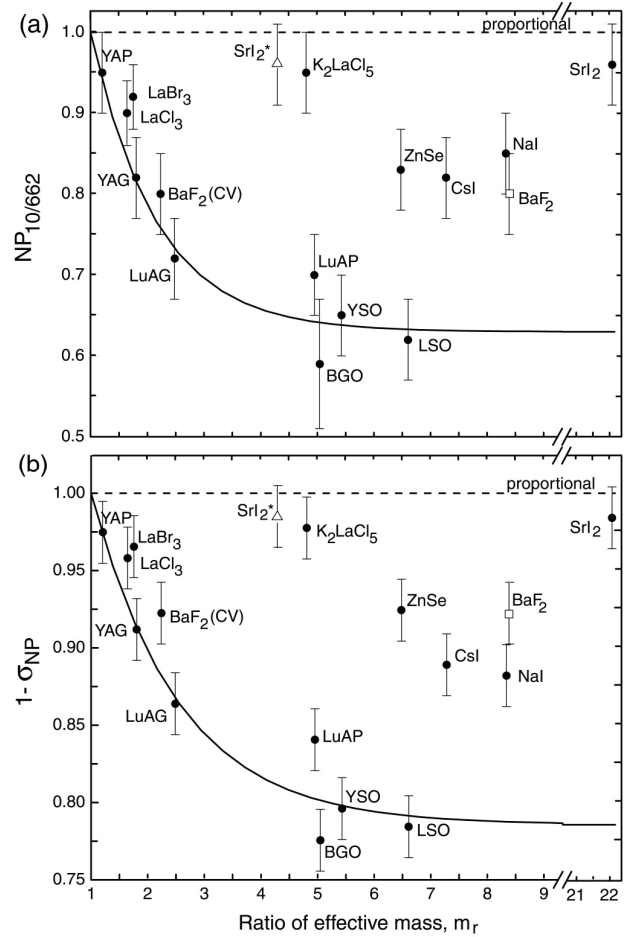

Fig. 2. Proportionality of photon response vs effective mass ratio. The plots show that for a group of materials, those with similar effective mass between electron and hole have superior proportionality response than otherwise. The fitting curve shows the trend. See text for a discussion on the BaF$_2$ and SrI$_2$ compounds.

The nonproportionality data relative to BaF$_2$ is for core-valence (CV) emission between F$^-$ 2p and Ba$^{2+}$ 5p orbitals. Instead of comparing the proportional behavior of this material to the ratio of the effective masses at the $\Gamma$-point (hollow square in Fig 2b), we have chosen to represent the effective mass of the electron and the hole calculated from the valence band minimum and the outermost core band maximum respectively (black circle in Fig 2b).

## IV. DISCUSSION

### A. Ambient Pressure

Figure 2 shows the effective mass ratio $m_r$ (defined in (3)) versus the proportionality of photon response: (a) $NP_{10/662}$ and (b) 1-$\sigma_{NP}$. In both plots, a group of materials, including YAP, LuAP, LSO and BGO, follows a consistent trend for which the proportionality decreases as the ratio of the effective masses increases. This trend is illustrated by fitting the data to a first-order exponential decay function with a constraint of passing the coordinate (1,1) representing a proportional system with electrons and holes having the same effective mass. The resulting curve is shown as solid line. A similar dependency, albeit weaker, is also observed in Fig 2b for a second category of materials including KLC, ZnSe, CsI and NaI.

It is known that scintillation process may be divided into three stages: the ionization via avalanche electron-electron scattering and the thermalization of electrons and holes via electron-phonon dissipation, further relaxation and formation of excitonic states in ionic materials and energy transport to the luminescence centers, and luminescence. From a comparison of the time scale of electron-electron relaxation ($10^{-15}$-$10^{-13}$ s), electron-phonon relaxation ($10^{-12}$-$10^{-11}$ s), and lifetime of excited luminescence centers ($>10^{-9}$ s), we may consider these processes as events following each other [8].

After the thermalization, the excited electrons (holes resp.) move down (up resp.) and fill the bottom (top resp.) of the conduction (valence resp.) band. The effective mass of charge carriers depends on the band curvatures. Different types of energy carriers such as excitons and electron and hole pairs may exist in a crystal. The idea that both type of excitations (exciton and free e/h) may be present in a scintillator and that their relative proportion affects the shape of the light yield response is adapted in a recent phenomenological model relating recombination rates and nonproportionality [17]. Assuming similar carriers' lifetime in a particular scintillator, the spatial distribution of free charge carriers is sensitive to the difference in their effective masses. This is not the case in excitonic type scintillators as the electrons and holes are coupled and migrate together as a whole. In the former case, when both free electrons and holes can be found in nearby regions of luminescence centers, the energy loss is decreased and the materials are then expected to have good proportionality. Experimental evidence that explicitly quantifies the relative proportion of free charge carriers to excitons has not been found in the literature. However, the scintillation mechanisms in YAP and LuAP [37, 38], YAG and LuAG [39, 40], YSO [41], LSO [ 42, 43], and BGO [44-46] have been shown to involve shallow and deep electron and hole traps, therefore suggesting the predominance of free carriers over excitons. The dependency observed in Fig 2b for these scintillator materials, indeed suggests that free electrons and holes are the main energy carriers.

On the other hand, excitonic luminescence is usually ascribed to alkali halide scintillators due to the strong exciton-phonon interactions which result in exciton or hole being self-trapped into a localized state with appreciable lattice distortion [47]. In LaBr$_3$ and LaCl$_3$, interesting interplay between free electrons and holes and self-trapped excitons involved in the scintillation is found [48]. The immediate capture of free electrons and holes by Ce$^{3+}$ leads to the temperature-independent component of the scintillation while the thermally activated self-trapped excitons (STE) is responsible for the temperature-dependent part. The migration mechanism of self-trapped holes (V$_k$ centers), free excitons (FE) and STE is different from that of free charge carriers. Furthermore, the lattice distortion induced by V$_k$ centers and excitons would result in different band dispersion from that of undistorted crystals modeled in our study. The weaker dependency observed between nonproportionality and the effective mass ratio, in the KLC, ZnSe, CsI, and NaI compounds, supports this idea as the mobility of individual charge carriers is no longer the main parameter to describe the energy transfer.

In the case of cross-luminescent materials, exemplified here by BaF$_2$, the radiative lifetime of the core-hole is relatively longer than the characteristic time for lattice vibrations ($10^{-9}$ s vs. $10^{-12}$ s respectively). This leads to the complete relaxation of the electronic system before radiative emission occurs (i.e. by thermalization of the core-hole to the top of the uppermost core-band) and relaxes the lattice near the core-hole (deformation). This results in the localization of the core-hole on the barium sites (self-trapping). This is why core-luminescence is treated as a radiative recombination of valence electrons with the localized core-holes. However, because some energy barrier exists between the free and the localized states of core-holes, they experience some mobility before their localization [49]. Other experimental data would be needed to confirm that cross-luminescent materials indeed behave as free charge carrier type scintillators.

Using the average effective mass of charge carriers defined by eq. 1 and 2, one notices that SrI$_2$ stands radically out of the previously described trends (black circle in Fig 2b). Understanding the reasons for this peculiar behavior would be of particular interest since this recently identified scintillator material is known, besides its high brightness, to be very proportional. Compared to the other materials studied in this paper, SrI$_2$ has a unique interlayered crystal structure (see Fig 3) which results in largely anisotropic band curvatures at the Γ-point (Table 2). The waviness of the iodine layers (in the X-Y direction) associated to the foliation of the structure (along the Z direction) very likely lead to a preferential direction for the energy migration. If one uses the effective masses of electrons and holes corresponding to the X-Γ-X direction ($m_r$ = 4.30) in place of the average of the effective masses ($m_r$ = 22.08), one finds that SrI$_2$ actually lies on the trend followed by excitonic type scintillator materials (triangle in Fig 2b). Even though the X-Γ-X direction has the smallest effective mass ratio compared to Y-Γ-Y and Z-Γ-Z, the lightest electron is found in the Z-Γ-Z direction. In addition, the CBM and VBM do not always occur at the same **k**-point (Table I). Therefore, more study would be needed to quantify the role played by the anisotropy of the crystal structure in the non-proportionality.

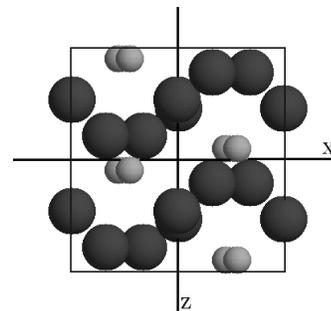

Fig. 3. Crystal structure of SrI$_2$ projected along the Y-direction. Iodine is represented by the large atoms.

TABLE II

CALCULATED EFFECTIVE MASSES OF ELECTRONS ($m_e$) AND HOLES ($m_h$) IN $SrI_2$ NEAR THE $\Gamma$-POINT.

| Directions in k-space | $m_e$ ($\Gamma$-point) [$m_0$] | $m_h$ ($\Gamma$-point) [$m_0$] | $m_r$ ($\Gamma$-point) |
|---|---|---|---|
| X-$\Gamma$-X | 0.303 | 1.304 | 4.304 |
| Y-$\Gamma$-Y | 0.301 | 5.699 | 18.93 |
| Z-$\Gamma$-Z | 0.259 | 12.076 | 46.62 |
| Average (eq. 1 and 2) | 0.288 | 6.360 | 22.08 |

### B. Influence of Pressure on Effective Masses

One way to alter band structure non-invasively is to modify the lattice constant by applying considerable pressure. Through our modeling, we have verified that, under pressure, it is possible to change the values of $m_e$ and $m_h$ of some scintillator materials while the crystal structure remains unmodified (except in the case of $BaF_2$ and $SrI_2$). $BaF_2$ undergoes phase transition from cubic (#225) to orthorhombic (#62) at about 3 GPa [50]. $SrI_2$ undergoes a phase transition within the orthorhombic group (from #61 to #62) at about 1.5 GPa [51]. Figure 4 shows the electronic structures of LuAP, $SrI_2$, and BGO at ambient pressure (left panel) and at 10, 2, and 6 GPa respectively (right panel). Figure 4(a) and 4(b) show that the "dip" at the CBM of LuAP flattens at high pressure indicating that the effective mass of electron increases. Meanwhile, there is no considerable change in the band dispersion of the VBM at $\Gamma$ point. For $SrI_2$ and BGO, the curvatures at the VBM changes while the CBM does not, as shown in Figure 4(c) to 4(f). In $SrI_2$, the VBM which occurs at the $\Gamma$ point becomes curved under pressure. The opposite phenomenon is found for BGO, the VBM which occurs at H point becomes flatter under pressure. Motivated by this observation, we systematically perform high throughput band structure calculations of all materials under study at 0, 0.5, 1, 2, 4, 6, 8, and 10 GPa. We extract the effective masses and calculate the $m_r$.

The evolution of effective mass ratio as a function of pressure is plotted in Figure 5. For clarity, the plots are presented in two groups: Figure 5(a) for NaI, CsI, and $SrI_2$, KLC, $LaBr_3$, $LaCl_3$, $BaF_2$ (CV), and ZnSe, and Figure 5(b) for YAP, YAG, LuAG, LuAP, YSO, BGO, and LSO. The curve for $LaBr_3$ (triangle) and $LaCl_3$ (circle) is similar and can barely be distinguished in Figure 4(a). Materials having $m_r$ generally decreasing with increasing pressure include NaI, CsI, $SrI_2$, ZnSe, KLC, YSO, LSO, and LuAP. For YAP, YAG, LuAG, $BaF_2$ (CV), $LaBr_3$, and $LaCl_3$ the behavior is relatively unchanged throughout the pressure range. A peculiar curve is obtained for BGO in which $m_r$ increases sharply at pressure from 0 to 6 GPa, and then decreases at higher pressure. Considering that the range of pressure that is accessible in experiments might be limited to $P < 1$ GPa, BGO offers an excellent case to further test the observed trend between effective mass ratio and nonproportionality.

### V. CONCLUSION

A correlation between the effective mass ratio of charge carriers and nonproportionality of photon response in aluminates, garnets, orthosilicates, oxides, $LaBr_3$ and $LaCl_3$ has been found. A similar trend was also observed for alkali halides, $K_2LaCl_5$, $BaF_2$, ZnSe. This suggests that the observed trend is due to free electrons and holes being the energy carriers to the luminescence centers and that the effective mass ratio between them is critical in understanding the nonproportionality of the response.

Since the parameter $m_r$ can be easily extracted from the band structures, a simple parameterization may lead to a high throughput materials design for proportional scintillators. Future experimental and theoretical work looking at the role of anisotropy in crystal structures and the effect of pressure on nonproportionality will help elucidate the observed correlation between the intrinsic resolution and the electronic band-structure of inorganic scintillator materials.

### ACKNOWLEDGMENT

We acknowledge Prof. Walter Harrison for constructive discussions. We thank the Teragrid Partnership (Texas Advanced Computing Center, TACC MCA-07S005) for computational support.

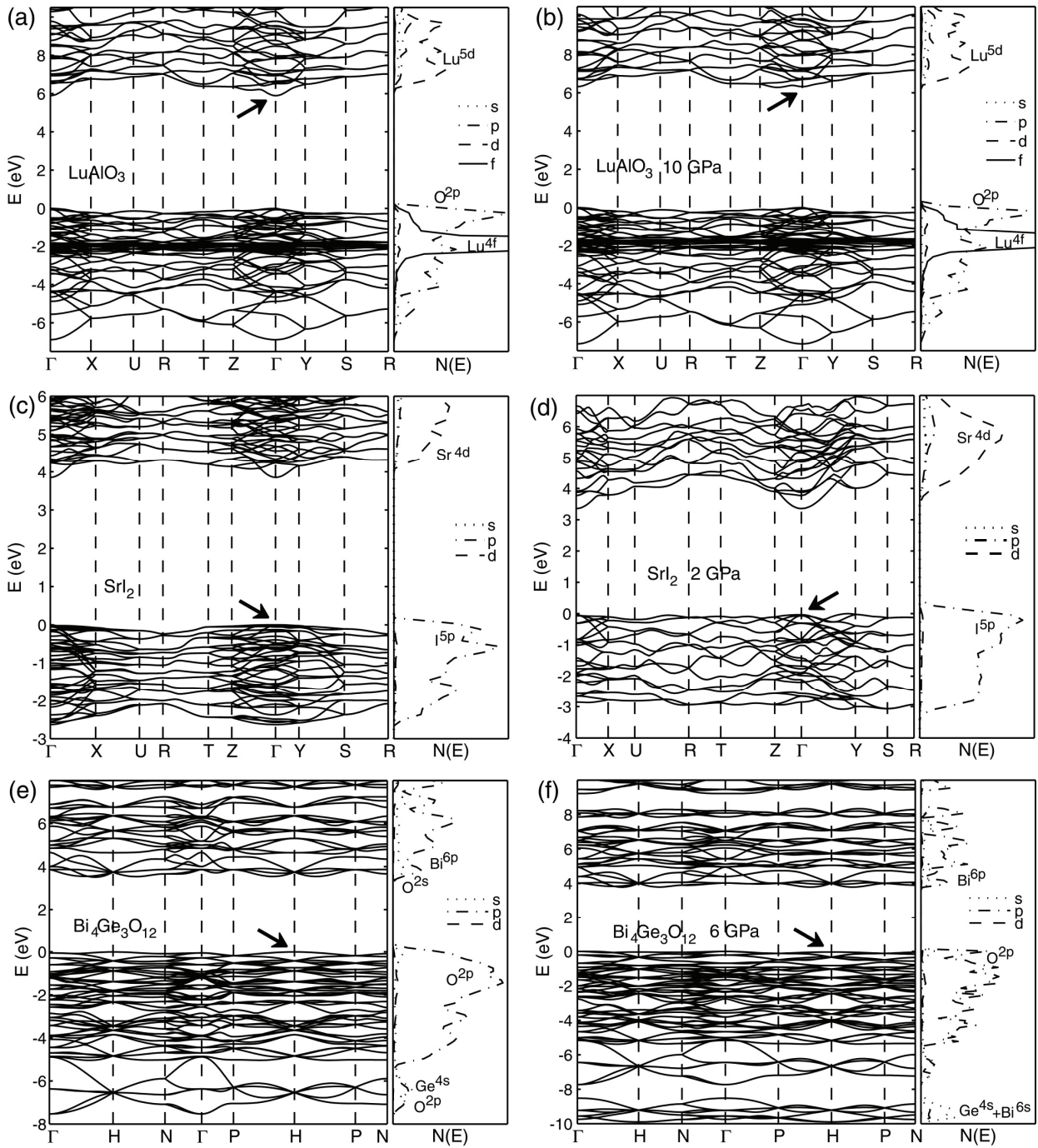

Fig.4. Calculated electronic structures of LuAlO₃, SrI₂, and BGO at ambient pressure (left panels) and at higher pressure (right panels). The scale is shifted so that the top of valence bands is at zero. The plots show the effect of pressure of the band dispersion as as indicated with arrows. Note that SrI₂ undergoes a phase transition within orthorhombic group from (#61) to (#62) at ∼ 1.5 GPa [51].

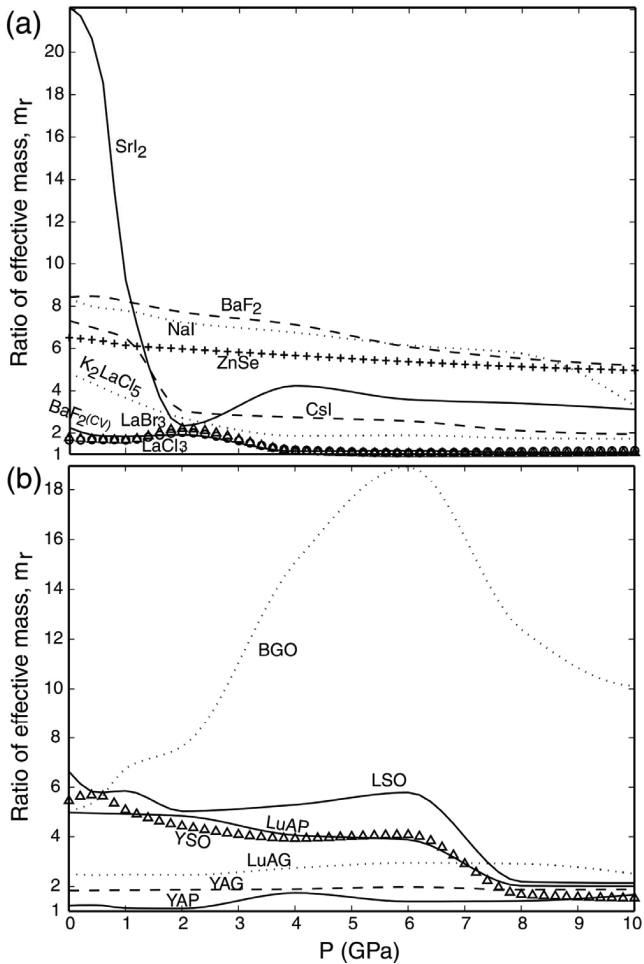

Fig. 5. Evolution of the calculated effective mass ratio, $m_r$, of selected scintillator crystals as a function of pressure up to 10 GPa.